\documentclass[a4paper,12pt,reqno,superscriptaddress,showkeys,nofootinbib]{revtex4}
\usepackage[centertags]{amsmath}
\usepackage{amsfonts}
\usepackage{amssymb}
\usepackage{amsthm}
\usepackage{newlfont}
\usepackage{stmaryrd}
\usepackage{mathrsfs}
\usepackage{euscript}
\usepackage{siunitx}
\usepackage{algorithm2e}
\usepackage{graphicx,subcaption}
\usepackage{graphicx,caption}
\usepackage{enumitem}
\usepackage{natbib}
\usepackage{graphicx}
\usepackage{xcolor}
\usepackage{floatrow}
\usepackage{caption}
\usepackage{hyperref}
\usepackage{hhline}
\usepackage{hyperref}
\usepackage{cleveref}
\usepackage{import}
\usepackage{verbatim} 

\theoremstyle{plain}
  \newtheorem{theorem}{Theorem}[section]

\theoremstyle{definition}

\theoremstyle{remark}

  \newtheorem{example}[theorem]{Example}

\DeclareCaptionJustification{justified}{\leftskip=0pt \rightskip=0pt \parfillskip=0pt plus 1fil}

\numberwithin{equation}{section}

\newcommand{\bbR}{{\mathbb R}}

\newcommand{\opunit}{\text{1}\kern-0.22em\text{l}}

\DeclareMathAlphabet{\mathpzc}{OT1}{pzc}{m}{it}

\newcommand{\fig}{Fig.\;}

\usepackage{floatrow}
\usepackage{caption}

\begin{document}
\title{Affine relationships between steady currents}
\author{Faezeh Khodabandehlou}
	\affiliation{Department of Physics and Astronomy, KU Leuven, Belgium}
		\email{faezeh.khodabandehlou@kuleuven.be}
\author{Christian Maes}
	\affiliation{Department of Physics and Astronomy, KU Leuven, Belgium}
 \author{Karel Neto\v{c}n\'{y}}
\affiliation{Institute of Physics, Czech Academy of Sciences, Prague, Czech Republic}
\begin{abstract}
Perturbing transition rates in a steady nonequilibrium system, {\it e.g.} modeled by a Markov jump process, causes a change in the local currents.  Their susceptibility is usually expressed via Green-Kubo relations or their nonequilibrium extensions.  However, we may also wish to directly express the mutual relation between currents.
Such a nonperturbative interrelation was discovered  by P.E.  Harunari {\it et al.} in  \cite{viv1} 
by applying algebraic graph theory showing the mutual linearity of currents over different edges in a graph.   We give a novel and shorter derivation of that current relationship where we express the current-current susceptibility as a difference in mean first-passage times.  It allows an extension to multiple currents, which remains affine but the relation is not additive.
\end{abstract}

\maketitle

\section{Introduction}
The maintenance of currents or fluxes is a key-feature of steady nonequilibrium systems.  It is of great importance there to understand the current susceptibility, {\it i.e.}, how it relates to (thermodynamic) forces or possible changes in architecture {\it etc}.  Around equilibrium, the result gets traditionally summarized by means of linear response or transport coefficients such as mobilities or conductivities.   That is part of standard linear response theory and goes under the name of Green-Kubo relations, \cite{green, kubo}.
The transport coefficients usually disclose a more microscopic specification of the system's susceptibility to driving.  Around equilibrium, they directly reveal the Onsager(-Casimir) reciprocity as a consequence of time-reversal invariance, \cite{onsa}.
Around nonequilibrium conditions, the extended Green-Kubo relations become more involved as they need to incorporate corrections to the equilibrium version  of the fluctuation-dissipation theorem; see {\it e.g.} \cite{resp,gaspard}.\\

Recently, an alternative to characterizing the current response has appeared, at least in the context of continuous-time Markov jump processes, {\it e.g.} for describing unimolecular chemical reactions in open networks.  Harunari, Dal Cengio, Lecomte and Polettini have proven that any two stationary currents are linearly related upon perturbations of a single edge's transition rates, arbitrarily far from equilibrium, \cite{viv1}.
They give explicit expressions for the current-current susceptibility in terms of the network geometry, \cite{viv2}.\\

The present paper continues that work by giving an elementary derivation, extension and interpretation of their relation.  We start by expressing how the stationary distribution gets modified when changing transition rates.  It constitutes a slight generalization of the setup in \cite{viv1}, where we work with multigraphs and hence allow for different transition channels over an edge. In that way, we can compare the stationary distributions when we remove or add channels.  From there we readily get the mutual linearity between steady currents in different multigraphs, which we generalize to an affine relation between multiple currents.\\  Interesting and also new is that the current-current susceptibility gets a physical interpretation in terms of first-passage times.  The connection between a response coefficient and first-passage times is original but natural as it expresses the causal structure between the directed motion of particles in different locations.\\
In the Appendix, we add a graphical expression for the current-current susceptibility, starting from the Kirchhoff formula for the steady currents. Such graphical representations clarify the relevant graph properties for the transport of current, and hence make it possible to micro-engineer or control how currents propagate through the network.\\  

We have in mind applications like the regulation of traffic flow in complex networks to the control of biological pathways subject to  boundary currents.  However, such applications may require the extension of our results to hypergraphs, which are not being considered here.\\

Plan of the paper: We start in the next section by giving the setup for comparing two Markov jump processes on a multigraph.  To comprehend the response in steady currents, it is natural to first understand how a perturbation affects the stationary distribution; that is done in Section \ref{sdis}.  The modification in stationary distribution when a transition rate is changed, is expressed via differences in mean first-passage times in the original process, Eq.\eqref{rho}.  Section \ref{affc} gives the corollary, yielding the affine current-current relation.  That gets generalized in Section \ref{mut} for multiple currents. The Example \ref{exthreeloops} emphasizes that the current-current relationship is not additive.
Appendix \ref{apkir} expresses the current-current susceptibility via the Kirchhoff formula.

 \section{Setup}
\begin{figure}[h]
      \centering
      \begin{subfigure}{0.49\textwidth}
         \centering
         \def\svgwidth{0.8\linewidth}        
        \includegraphics[scale = 0.73]{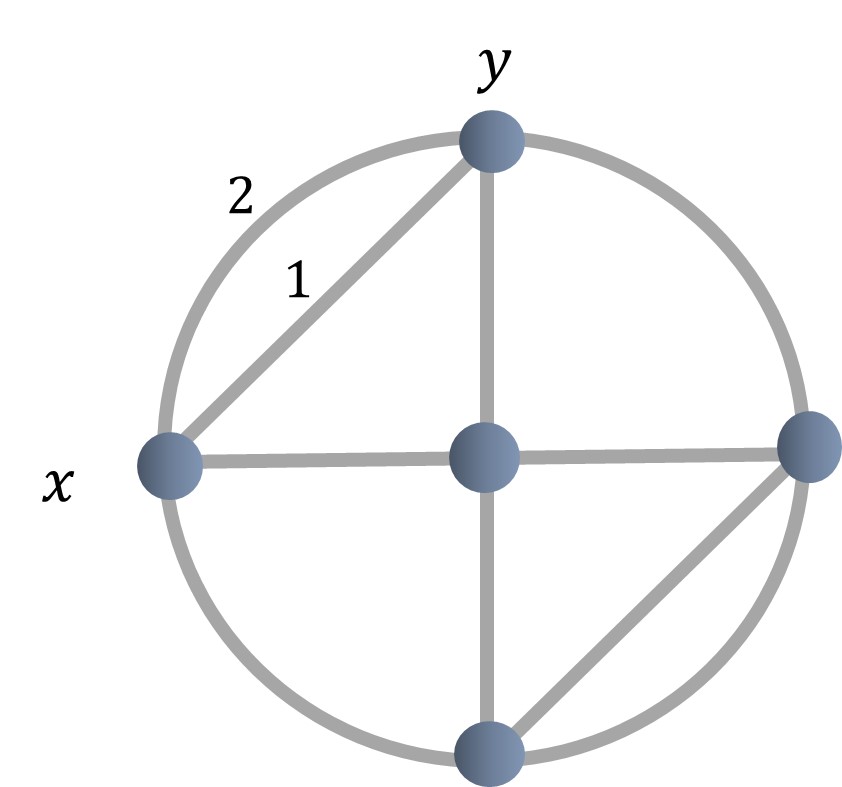}
        \caption{\small{}}
     \end{subfigure}
     \begin{subfigure}{0.49\textwidth}
         \centering
         \def\svgwidth{0.8\linewidth}        
   \includegraphics[scale = 0.73]{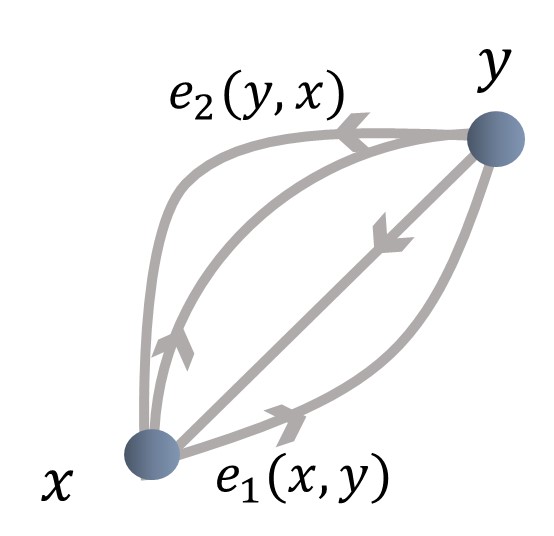}
           \caption{\small{}}
     \end{subfigure}
   \caption{\small{The system is modeled as a random walker on a multigraph (a), where each connection $x\leftrightarrow y$ may have different channels (b).
   }} \label{mg}
\end{figure}

The nonequilibrium system  has a  finite number of states $x,y,z,\ldots \in \cal V$, each for instance representing a configuration of spins, energies or molecules in the system under consideration. To understand the dynamics, we imagine a random walker hopping between those states.  The hopping may proceed through a number of possible channels. That Markovian stochastic dynamics is then mathematically specified by transition rates over the various channels that connect any two states.\\
More specifically, for the $i$th channel $x \overset{i}{\leftrightarrow}y$, let $k_i(x,y)$ and $k_i(y,x)$ denote the transition rates. In that way, we build a directed multigraph $G(\cal V, \cal E)$  where $\cal E$ is the set of all directed edges $e_i(x,y)$ between vertices $x$ and  $y$ where $k_i(x,y) >0$; see Fig.~\ref{mg}.\\

The above defines the Markov jump process with rates $k(x,y) =\sum_i k_i(x,y)$ and with forward generator $L^\dagger$ having matrix elements 
\begin{align}
    [L^\dagger]_{xy}&=k(y,x) \quad x\not=y, \qquad [L^\dagger]_{xx}=-\sum_{y}k(x,y)
\end{align}
Assuming the irreducibility of the graph, the unique stationary distribution is denoted by $\rho$,  normalized $\sum_x \rho(x)= 1$, with probabilities $\rho(x)>0$, and verifies the stationary Master equation $L^\dagger\rho=0$, for all  $x \in \cal V$,
\begin{align}\label{steadyM}
  \sum_y[k(y,x)\rho(y)-k(x,y)\rho(x)]=0
\end{align}
The steady current over the edge
$e_i(x,y)$ is defined as 
\begin{equation}
j_i(x,y)=k_i(x,y)\rho(x)-k_i(y,x)\rho(y)
\end{equation}
The sum $j(x,y)=\sum_ij_i(x,y)$ enters the steady Master equation \eqref{steadyM} as
\begin{equation}\label{steadyMa}
 L^\dagger\rho (x)=-\sum_yj(x,y)=0
\end{equation}

In what follows, the above notation refers to the situation {\it after} a change in transition rates. 
The original process is a Markov jump process on the same multigraph with forward generator $L_0^\dagger$ and (total) rates $k^0(x,y)$.  The change in those rates defines the forward generator $L^\dagger$, with (total transition) rates 
\begin{equation}\label{add}
k(x,y) = k^0(x,y) + \Delta(x,y)
\end{equation}
for $x\rightarrow y$.  The $\Delta(x,y)$  determine the change of transition rates.   In some cases, we can use that change to really change the channel structure.  The simplest to imagine is a single graph and to focus on one specific pair  $\{a,b\}$ of states: by taking $k^0(a,b)=0=k^0(b,a),\quad k(a,b)=\Delta(a,b),  k(b,a)=\Delta(b,a)$ we truly add an edge $a\leftrightarrow b$ with transition rates $\Delta(a,b), \Delta(b,a)$.\\

 Similarly and more generally, we detail $\Delta(x,y) = \sum_i \Delta_i(x,y)$ in \eqref{add} by writing per channel
\begin{equation}\label{adds}
k_i(x,y) = k^0_i(x,y) + \Delta_i(x,y)
\end{equation}
where we are especially interested in the case where $k_i^0(x,y)=0,  \Delta_i(x,y)>0$  (adding a channel for $x\rightarrow y$) and in the case where $\Delta_i(x,y)= -k_i^0(x,y)$  (removing a channel for $x\rightarrow y$).  In that sense, the change \eqref{adds} modifies the channel structure; it is like changing the set of edges $\cal E_0 \rightarrow \cal E$ over which there is a nonzero transition rate.  The set of states $\cal V_0=\cal V$ is never changing.   At any rate, mathematically it is most convenient to not change the graph geometry at all and to think of comparing two (or even more than two) Markov processes on the same graph.  We always assume that irreducibility is preserved.

\section{Change in stationary distribution}\label{sdis}

We wish to express the new stationary distribution $\rho$ in terms of the original (unperturbed) $\rho^0$ (where we always use sub- or superscript `$0$' to indicate the process with generator $L_0$ and rates $k^0(x,y)$).  In other words, how does the stationary distribution change $\rho^0\rightarrow \rho$ if we add channels or, more generally, change the transition rate over a channel as in \eqref{add}--\eqref{adds}?\\

Let  $\tau^0(z,x)$ be the mean first-passage time  to reach state $x$ when started from $z$, for the original process with rates $k^0(u,v)$, \cite{redner}.  We prove next that\footnote{The equation \eqref{rho} is the same as relation (44) in  the paper by Harvey \textit{et al.} \cite{timur}. We thank Timur Aslyamov for pointing that out, after reading our paper.} 
\begin{align} \label{rho}
    \rho(x) &=
    \rho^0(x)\left( 1+\sum_{z,z'}\tau^0(z,x)\,j_\Delta(z,z')\right)\\
        &=\rho^0(x)\left( 1+\frac 1{2}\sum_{z,z'}[\tau^0(z,x)-\tau^0(z',x)]\,j_\Delta(z,z')\right)\notag
\end{align}
where 
\begin{equation}\label{dj}
j_\Delta(z,z') = \Delta(z,z')\rho(z) - \Delta(z',z)\rho(z')
\end{equation}
Relation \eqref{rho} gives the modification in stationary distribution at $x$  where the change around $(z,z')$ is ``transported'' via the difference in mean first-passage times $\tau^0(z,x) - \tau^0(z',x)$ in the original process.\\   

Write $L^\dagger=L^\dagger_0+L^\dagger_\Delta$ and  put $u=L^\dagger_\Delta \rho$.  Then, $u(z)=-\sum_{z'}j_\Delta(z,z')$.  
Moreover,  $L^\dagger\rho=0$ (by stationarity) yields $L^\dagger_0 \rho=-u$, which has a solution because  $\sum_z u(z)=0$. Combined with $L^\dagger_0\rho^0=0$, we get 
\begin{equation}\label{l1}
L_0^\dagger(\rho-\rho^0)(z)=-u(z) = \sum_{z'}j_\Delta(z,z')
\end{equation}
Therefore, with backward generator $L_0$ and $ L_0 \tau^0(z,x) =\sum_{z'} k^0(z,z')[ \tau^0(z',x)-\tau^0(z,x)]$,
\begin{equation}\label{1s}
    \sum_z L_0 \tau^0(z,x) \,(\rho-\rho^0)(z) =\sum_z \tau^0(z,x) \, \, L^\dagger_0(\rho-\rho^0)(z)
    =\sum_{z,z'}\tau^0(z,x)j_\Delta(z,z')
\end{equation}
On the other hand,
the mean-first passage times  solve the equation 
 \begin{equation}\label{mf}
     L_0 \tau^0(z,x) =\frac{\delta_{xz}}{\rho^0(x)}-1 \quad \forall z\not=x, \qquad \tau^0(x,x)=0
 \end{equation}
 see {\it e.g.} \cite{pois}.  From multiplying \eqref{mf} with $(\rho-\rho^0)(z)$ and summing over $z$, \eqref{1s} must equal
\begin{equation}\label{2s}
\sum_z \big(\frac{\delta_{xz}}{\rho^0(x)}-1\big)\, (\rho-\rho^0)(z)= \frac{(\rho-\rho^0)(x)}{\rho^0(x)}.
\end{equation}
Hence,  \eqref{rho} follows from the right-hand sides being equal in \eqref{1s} = \eqref{2s}.

\section{Affine current relation}\label{affc}
We wish to understand how a current over an edge $e_i(x,y)$ is modified by changes (only) in the rates or channel-structure in a different location around edges $(z,z')$.
Let us therefore take a channel where $\Delta_i(x,y)=0 =\Delta_i(y,x)$ in \eqref{adds}.  Then, the corresponding channel current is
 \begin{align*}
    j_i(x,y)&=\rho(x)k_i^0(x,y)-\rho(y)k_i^0(y,x)\notag\\
    &=  j_i^0(x,y) +\sum_{z,z'}
\left(\rho^0(x) k_i^0(x,y)   \tau^0(z,x) - 
    \rho^0(y) k_i^0(y,x)    \tau^0(z,y)\right)\,j_\Delta(z,z') \\
    &=  j_i^0(x,y) + \sum_{z,z'}
    \lambda^0_i(zz',(x,y))\,j_\Delta(z,z')  
 \end{align*}
 where
we use \eqref{rho} in the second line and  \begin{equation}\label{lam}
  \lambda^0_i(zz',(x,y)) = \frac 1{2} \left( \rho^0(x) k_i^0(x,y)   [\tau^0(z,x)-\tau^0(z',x)] - 
    \rho^0(y) k^0_i(y,x)   [\tau^0(z,y)-\tau^0(z',y)] \right)
    \end{equation}
  does not depend on the $\Delta(z,z')$.\\
We conclude with our main result that  $\Delta_i(x,y)=0=\Delta_i(y,x)$ implies
 \begin{equation}\label{mr}
    j_i(x,y)=  j_i^0(x,y) + \sum_{(z,z')}
    \lambda^0_i(zz',(x,y))\,j_\Delta(z,z')  
 \end{equation}
 where the current-current susceptibilities $\lambda^0_i(zz',(x,y))$ are entirely expressed in the original (unperturbed) process, and from \eqref{dj},
\begin{align}
j_\Delta(z,z') &= \Delta(z,z')\rho(z) - \Delta(z',z)\rho(z')\notag\\
&= j(z,z') -  k^0(z,z')\rho(z) + k^0(z',z)\rho(z')\\
&= j(z,z') - j^0(z,z') -  \big(k^0(z,z')[\rho(z)-\rho^0(z)] - k^0(z',z)[\rho(z')-\rho^0(z')]\big),\notag
\end{align}
where $j^0(z,z')= \sum_i j^0_i(z,z') = k^0(z,z')\rho^0(z) - k^0(z',z)\rho^0(z')$.\\
We emphasize that $j_\Delta(z,z')$ is the (total) stationary current over $z\rightarrow z'$ when $k(z,z') = \Delta(z,z'), k(z',z) = \Delta(z',z)$ (or $k^0(z,z') = k^0(z',z) = 0$), and the stationary probabilities are $\rho(z),\rho(z')$.

\section{Mutual linearity of steady currents}\label{mut}

We start by returning to the simplest case of a single graph where an edge between vertices $a$ and $b$ is added, \cite{viv1}.  We put $k^0(a,b) = 0=k^0(b,a)$, so that $j_\Delta(a,b) = j(a,b)$.  From \eqref{rho} we find
 \begin{equation}\label{cu}
    \rho(x)= \rho^0(x)+\rho^0(x) [\tau^0(a,x)-\tau^0(b,x)]\,j(a,b)
\end{equation}
and, from \eqref{lam},  
\begin{equation}\label{orig}
 j(x,y) =  j^0(x,y) + 
    \lambda^0(ab,(x,y))\,j(a,b)      
\end{equation}
for
 \begin{align}\label{lambda1}
  \lambda^0(ab,(x,y)) =
k^0(x,y)\rho^0(x) [\tau^0(a,x)-\tau^0(b,x)]-k^0(y,x)\rho^0(y) [\tau^0(a,y)-\tau^0(b,y)]
\end{align}
and for any $(x,y)\not=(a,b), (b,a)$, the rates are $k^0(x,y)=k(x,y)$.\\

In \eqref{lam}, the current-current susceptibility is expressed in terms of differences of mean first-passage times.  However, there are still other representations.  (A graphical tree-representation is the subject of Appendix \ref{apkir}.)\\
Consider the quasipotential \( V^0_{xy}  = V^0 \) defined as the mean excess integrated current through a fixed edge $(x,y)$ (with $(x,y)\neq (a,b)$) when the original process is started from \( z \):
\begin{align}
V^0_{xy}(z) = \lim_{t \to \infty} \left[ \langle N_{xy}(t)\rangle_z^0 - \langle N_{xy}(t)\rangle_{\text{steady}}^0 \right]
\end{align}
where \( N_{xy}(t) \) increments by \( +1 \) for every transition \( x \to y \), and decrements by \( -1 \) for \( y \to x \) up to time \( t \). We show next the following alternative to \eqref{orig}: 
\begin{equation}\label{val}
j(x,y) = j^0(x,y) + (V^0_{xy}(b) - V^0_{xy}(a))\, j(a.b)
\end{equation}
By a standard argument, the quasipotential $V^0 =  V^0_{xy}$ satisfies the Poisson equation, \cite{pois}:
\begin{equation}
L^0 V^0\,(z) = j^0(x,y)- n(z), \quad  \qquad \sum_z \rho^0(z)\, V^0(z) = 0
\end{equation}
with $n(z) = n_{xy}(z)$ the expected instantaneous current along the edge $( x, y )$ at the time when the state is $z$:
\begin{align}
n(z) = \lim_{t \to 0^+} \frac{\langle N_{xy}(t)\rangle_z^0}{t} =
\begin{cases} 
k^0(x,y) & \text{if } z = x \\
-k^0(y,x) & \text{if } z = y \\
0 & \text{otherwise}
\end{cases}
\end{align}
Note that its steady-state expectation equals $\langle n\rangle^0 = j^0(x,y)$.\\
Put $\nu(x)=\rho^0(x) [\tau^0(a,x)-\tau^0(b,x)]$.  We know from \eqref{cu} that
$\nu(x) j(a,b) = \rho(x) - \rho^0(x)$, and from \eqref{l1} that
\[
L_0^\dagger(\rho-\rho^0)(z)= (\delta_{za} - \delta_{zb})\,j(a,b)
\]
Hence, $L_0^\dagger\nu (z) = \delta_{za}-\delta_{zb}$.  Therefore,
\begin{align}
    V_{xy}^0(a) - V_{xy}^0(b) =& \sum_zV^0(z) (\delta_{za}-\delta_{zb})  = \sum_z V^0_{xy}(z) \,L_0^\dagger\nu (z)= \sum_z L^0 V_{xy}^0 (z) \,\nu(z)\\
= &\sum_z (j^0(x,y) - n(z))\,\nu(z) = -\nu(x)k^0(x,y) + \nu(y) k^0(y,x) = \lambda^0(ab,(x,y))\notag
\end{align}
where the last equality uses \eqref{lambda1} and hence delivers \eqref{val}.\\

We can still generalize the above via \eqref{lam} by taking multiple edges $\{a,b\}$ where we put  $k^0(a,b) = 0=k^0(b,a)$ so that $j_\Delta(a,b) = j(a,b)$.  One should be careful however to use for the mean first-passage times $\tau^0(a,x)$ etc. the dynamics where indeed all  $k^0(a',b') = 0=k^0(b',a')$.  In other words, the result is different from taking  the sum of \eqref{orig}: the linear current relationship is not additive.  We illustrate that in the next example.

\begin{example}\label{exthreeloops}
Consider the graph given in \fig\ref{threeloops}. The vertices are labeled with the numbers from 1 to 9. 
\begin{figure}[H]
    \centering
    \includegraphics[scale=0.5]{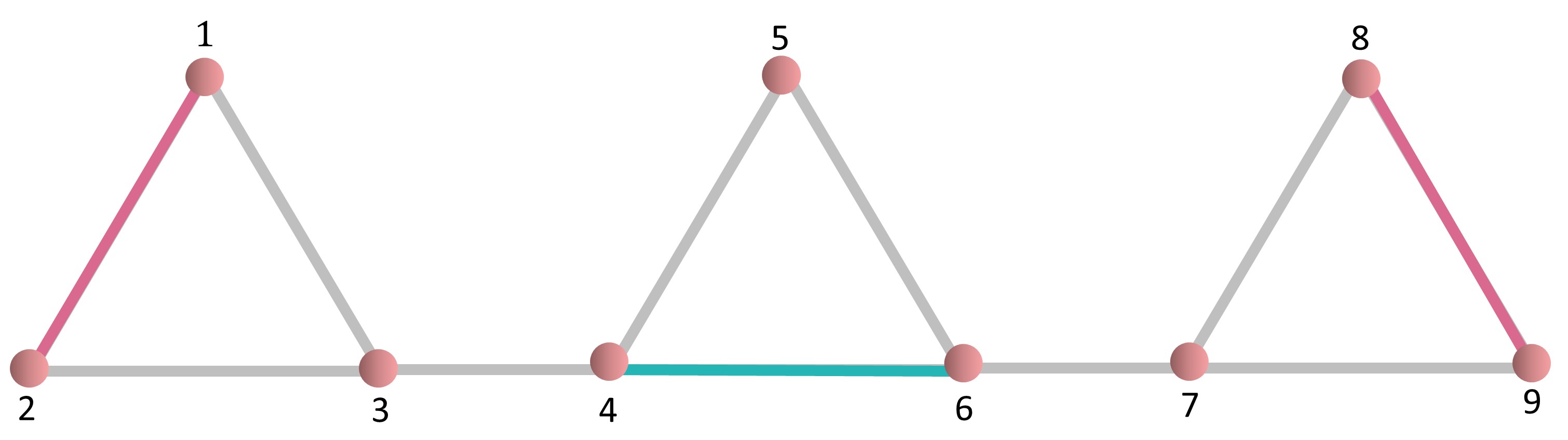}
    \caption{\small{The graph of Example \ref{exthreeloops}. The example illustrates the relationship between the current  over the edge $(4,6)$ with currents over the edges $(1,2)$ and $(8,9)$ and its nonadditivity.}}
    \label{threeloops}
\end{figure}
The transition rates are taken as (and putting $k_{uv}=k(u,v)$)
\begin{align}\label{rateex}
   & k_{12}=k_{23}=k_{31}=k_{78}=k_{89}=k_{97}= k_{46}=k_{65}=k_{54}=2\notag\\
& k_{21}=k_{13}=k_{32}=k_{98}=k_{87}=k_{79}= k_{46}=k_{65}=k_{54}=k_{15}=k_{51}=1\notag\\
   & k_{67}=k_{76}=k_{34}=k_{46}=1
    \end{align}
For this values of rates $j(4,6)=\frac{1}{9}$. Let $G^0_1$,  $G^0_2$ and  $G^0_3$ represent the graphs where, respectively in \fig\ref{threeloops}:   only the channel  $1 \leftrightarrow 2$,  only the channel $8 \leftrightarrow 9$, and finally, both of the aforementioned channels are removed. Notice that in this example, the subscripts $1, 2 $ and  $3$ do not represent the channel number since the graph in \fig \ref{threeloops} is a single graph.\\

Using \eqref{mr},
\begin{align}
    j(4,6)=j^0_3(4,6)+\lambda^0_3(12,(4,6))j(1,2)+ \lambda^0_3(89,(4,6))j(8,9),
\end{align}
The mean first-passage times are obtained from the Poisson equation \eqref{mf}.  Plugging them into \eqref{lam} we find
 \begin{align}
   j(4,6)&=j_1^0(4,6)+\lambda_1^0(12,(4,6))j(1,2)=\frac{2}{19}+\frac{1}{19}j(1,2) \label{j1}\\
   j(4,6)&=j_2^0(4,6)+\lambda_2^0(89,(4,6))j(8,9)=\frac{2}{19}+\frac{1}{19}j(8,9)\label{j2}\\
   j(4,6)&=j^0_3(4,6)+\lambda^0_3(12,(4,6))j(1,2)+ \lambda^0_3(89,(4,6))j(8,9)\notag\\
   &\qquad=\frac{1}{10} +\frac{1}{20}j(1,2)+\frac{1}{20}j(8,9)\label{j3}
 \end{align}
 It is important to note that the relation \eqref{j3} is not just obtained by adding \eqref{j1} and \eqref{j2}. That illustrates the nonadditivity of the current relationship.
\end{example}

\section{Conclusion}
We have given a shorter and transparent derivation of the mutual linearity between currents in a nonequilibrium Markov jump process on a graph, \cite{viv1}.
Moreover, we have been able to express the current-current susceptibility in terms of differences of mean first--passage times or quasipotentials associated to an excess current. The connection enables a new physical interpretation and offers a separate road for computing the relationship.  That structure is preserved in the affine relation between multiple currents and holds in the multigraph context.

\vspace{1cm}

\noindent {\bf Acknowledgment}:  We thank Vivien Lecomte for useful discussions and remarks, and for motivating us to meditate on the results in \cite{viv1} and to write our own version and extensions.

\appendix
\section{Tree-representation of the current-current susceptibility}\label{apkir}
Applying Kirchhoff's formula, we give a graphical representation of the `susceptibility' $\lambda^0$ appearing in the affine relation and thus giving an alternative to the expression \eqref{orig}. In this section, we assign an edge $\{x,y\}$ between every two vertices $x$ and $y$ iff  $k(x,y)\not=0$ and $k(y,x)\not=0$.  Hence, the edge has two possible directions: $(x,y)$ and $(y,x)$.\\

 Kirchhoff's formula gives a graphical representation of stationary distribution  $\rho$ in \eqref{steadyM},
\begin{align}\label{kir}
    \rho(x)=\frac{w(x)}{W},
\end{align}
where 
\[w(x)=\sum_\cal T w(\cal T_x), \quad W=\sum_yw(y).\]
The first sum is over all spanning trees \cite{intro}. A spanning tree on the connected graph $G$ connects all vertices where the number of edges in the spanning tree is $|\cal V|-1$. A single vertex is considered a tree. $\cal T$ denotes a spanning tree. If all edges in the spanning tree are directed toward a specific vertex, the tree is called a rooted spanning tree. $\cal T_x$ denotes a spanning tree rooted in $x$.  \\
The weight $k(x,y)$ is assigned to the edge $(x,y)$. $w(\cal T_x)$ denotes the weight of $\cal T_x$, equal to the product of all weights (rates assigned to edges) in the tree.\\

According to the Kirchhoff formula \eqref{kir}, the steady current over a fixed $(x,y)$ is  
\begin{align}
    j(x,y)=\frac{w(j_{xy})}{W}, \quad w(j_{xy})=k(x,y)w(x)-k(y,x)w(y).
\end{align}
\\
For every other edge $(a,b)$, which is not a bridge, we have the affine current--current relation as in \eqref{orig}:
    \begin{align}\label{jj}
        j(x,y)=j^0(x,y)+\lambda^0(ab,(x,y))\, j(a,b),
    \end{align}
   where  $j^0(x,y)$ and $\lambda^0(ab,(x,y))$ neither depend on  $k(a,b)$ nor $k(b,a)$, and 
   \begin{align}\label{lambda2}
       j^0(x,y)=\frac{w^0(j_{xy})}{W^0} \qquad \lambda^0(ab,(x,y))=\frac{w^{(a,b)}(j_{xy})-j^0(x,y) W^{(a,b)}}{k(a,b)w^0(a)},
   \end{align}
   where  $\lambda^0(ab,(x,y))=-\lambda^0(ba,(x,y))$. 
We use the superscript $ ^0$ to denote the weight of the subgraphs that do not have the directed edges $(a,b)$ and $(b,a)$. The superscript $^{(a,b)}$ is used for the weight of subgraphs including $(a,b)$, and the superscript $^{(b,a)}$ denotes the weight of subgraphs including $(b,a)$. \\
Notice that  $j^0(x,y)$ represents the steady current over the edge $(x,y)$ where $k(a,b)=k(b,a)=0$. Although $j^0(x,y)$ and  $\lambda^0(ab,(x,y))$ are independent of $k(a,b)$ and $k(b,a)$, they do depend on the full geometry. 
The notation in \eqref{jj}--\eqref{lambda2} follows
 \begin{align*}
     w^0(j_{xy}) =&\,  w^0(x)k(x,y)-w^0(y)k(y, x)\\
     =& \sum_{\cal T \not \ni \{x,y\}}[w(\cal T_{x})k(x,y)-w(\cal T_{y})k(y, x)],\\
      w^{(a,b)}(j_{xy}) =& \, w^{(a,b)}(x)k(x,y)-w^{(a,b)}(y)k(y, x)\\
      =& \sum_{\cal T_{x} \ni (a,b)}w(\cal T_{x})k(x,y)-\sum_{\cal T_{y} \ni (a,b)}w(\cal T_{y})k(y, x),\\
       w^{(b,a)}(j_{xy}) =& \, w^{(b,a)}(x)k(x,y)-w^{(b,a)}(y)k(y, x).
 \end{align*}
Similarly, put  $W^0=\sum_{\cal T \not \ni \{x,y\}}\sum_u w(\cal T_u)$ which is the weight of all rooted spanning trees that do not contain $(a,b)$ nor $(b,a)$, and 
 \begin{align*}
W^{(a,b)}=\sum_u\sum_{\cal T_u \ni(a,b)} w(\cal T_u); \qquad W^{(b,a)}=\sum_u\sum_{\cal T_u \ni(b,a)} w(\cal T_u).
 \end{align*}
It is clear that $w(j_{xy})=w^0(j_{xy})+w^{(a,b)}(j_{xy})+w^{(b,a)}(j_{xy})$, and $W=W^0+W^{(a,b)}+W^{(b,a)}$.

\begin{example}
To clarify the notation, consider the graph $G$ given in \fig~\ref{ex}.
\begin{figure}[H]
    \centering
    \includegraphics[scale=0.8]{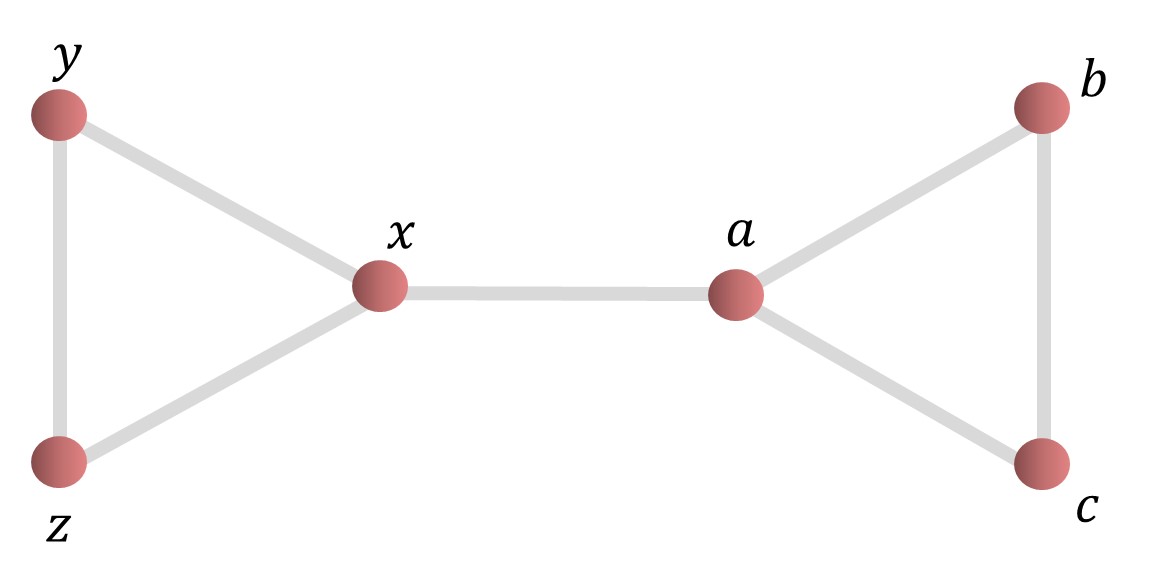}
    \caption{\small{Two loops connected via one edge.}}\label{ex}
    \label{ex}
\end{figure}

\begin{align*}
    &w^0(j_{xy})= [k_{xy}k_{yz}k_{zx}- k_{yx}k_{xz}k_{zy}]k_{bc}k_{ca}k_{ax},\\
    &w^{(a,b)}(j_{xy})= 0,\\
    &w^0(x)=[k_{zy}k_{yx}+k_{yx}k_{zx}+k_{yz}k_{zx}]k_{ax}k_{bc}k_{ca},\\
    &W^{(a,b)}=[k_{zy}k_{yx}+k_{yx}k_{zx}+k_{yz}k_{zx}]k_{xa}k_{ab} (k_{cb}+k_{ca}+k_{bc})
\end{align*}
and after some simplifications,
\begin{align*}
j^0(x,y)&=\frac{[k_{xy}k_{yz}k_{zx}- k_{yx}k_{xz}k_{zy}]k_{bc}k_{ca}k_{ax}}{W^0},\\
    \lambda^0(ab,(x,y))&=-\frac{k_{ax} \left(k_{bc}+k_{ca}+k_{cb}\right) \left(k_{yz} k_{zx} k_{xy}-k_{yx} k_{zy} k_{xz}\right)}{W^0},
\end{align*}
where $W^0=\sum_{u\in \cal V}w^0(u)$.
\end{example}

We end with the proof of \eqref{jj}.\\
Fix the edge $(x,y)$.  We can write  $j(x,y)=j^0(x,y)+\lambda^0 (ab,(x,y))\, j(a,b)$ and $j(x,y)=j^0(x,y)+\lambda^0(ba,(x,y))\, j(b,a)$. Using $j(a,b)=-j(b,a)$, then  $\lambda^0 (ab,(x,y))=-\lambda^0(ba,(x,y))$.\\
First, notice that the steady current over the edge $(a,b)$ is 
\begin{align*}
    j(a,b)=\frac{w(j_{ab})}{W}&=\frac{1}{W}\sum_\cal T[w(\cal T_a)k(a,b)-w(\cal T_b)k(b,a)]\\
    &=\frac{1}{W}\sum_{\cal T\not \ni \{a,b\}}[w(\cal T_a)k(a,b)-w(\cal T_b)k(b,a)]\\
    &=\frac{1}{W}[w^0(a)k(a,b)-w^0(b)k(b,a)]
\end{align*}
because, for every spanning tree $\cal T$ including $\{a,b\}$, $w(\cal T_a)k(a,b)=w(\cal T_b)k(b,a)$.\\

For an arbitrary edge $(a,b)$ which is not a bridge, 
\begin{align*}
   j^0(x,y)+\lambda^0 (ab,(x,y)) j(a,b)&= j^0(x,y)+ \lambda^0  (ab,(x,y))(\frac{w^0(a)k(a,b)-w^0(b)k(b,a)}{W})\\
    &=j^0(x,y)+ (\frac{w^{(a,b)}(j_{xy})-j^0(x,y) W^{(a,b)}}{k(a,b)w^0(a)})(\frac{w^0(a)k(a,b)}{W})\\
    &\quad+(\frac{w^{(b,a)}(j_{xy})-j^0(x,y) W^{(b,a)}}{k(b,a)w^0(b)})(\frac{w^0(b)k(b,a)}{W})\\
    &=j^0(x,y)+ (\frac{w^{(a,b)}(j_{xy})-j^0(x,y) W^{(a,b)}}{W})+\frac{w^0(j_{xy})}{W}\\
    &\quad \qquad \quad +(\frac{w^{(b,a)}(j_{xy})-j^0(x,y)W^{(b,a)}}{W})-\frac{w^0(j_{xy})}{W}
\end{align*}
using $j(x,y)=\frac{w(j_{xy})}{W}=\frac{w^0(j_{xy})+w^{(a,b)}(j_{xy})+w^{(b,a)}(j_{xy})}{W}$, then
\begin{align*}
    j^0(x,y)+\lambda^0  (ab,(x,y))j(a,b)&=j^0(x,y)-j^0(x,y)(\frac{W^{(a,b)}+W^{(b,a)}}{W})\\
    & \qquad-\frac{w^0(j_{xy})}{W}+j(x,y)\\
    &=j^0(x,y)(\frac{W-(W^{(a,b)}+W^{(b,a)})}{W})-\frac{w^0(j_{xy})}{W}+j(x,y)\\
    &=\frac{w^0(j_{xy})}{W^0}\frac{W^0}{W}-\frac{w^0(j_{xy})}{W}+j(x,y)=j(x,y)
\end{align*}
which ends the proof of \eqref{jj}.

\bibliographystyle{unsrt}  
\bibliography{chr}
\onecolumngrid

\end{document}